\documentclass[aps,prd,twocolumn,showpacs,showkeys,final,floatfix,superscriptaddres,s
preprintnumbers,nofootinbib]{revtex4}
\usepackage{graphicx,amssymb,url}
\usepackage{amsmath}
\usepackage{showkeys}

\begin{document}
\title{Accretion with back reaction}
\author{Vyacheslav I. Dokuchaev}
\author{Yury N. Eroshenko}
\affiliation{Institute for Nuclear Research of the Russian Academy of Sciences \\ 60th October
Anniversary Prospect 7a, 117312 Moscow, Russia}

\begin{abstract}
We calculate analytically a back reaction of the stationary spherical accretion flow near the
event horizon and near the inner Cauchy horizon of the charged black hole. It is shown that
corresponding back reaction corrections to the black hole metric depend only on the fluid
accretion rate and diverge in the case of extremely charged black hole. In result, the test
fluid approximation for stationary accretion is violated for extreme black holes. This behavior
of the accreting black hole is in accordance with the third law of black hole thermodynamics,
forbidding the practical attainability of the extreme state.
\end{abstract}


\maketitle

\section{Introduction}

The matter and fields, inflowing into a rotating or charged black hole, generate the drastic
singular phenomena at the inner Cauchy horizon: the infinite blue-shift and mass inflation (see
e.\,g. \cite{Penrose68,SimPen,Gursel79,Gursel79b,Novikov80,Hiscock81,ChanHar82,PoisIs90,
Ori91,Gnedin93,Bonanno94,BurkoOri95,BradySmith95,HodPiran97,Burko97,BradyMM98,NolWat02,HKN05,
HamAvel10}). We describe here a back reaction of the stationary spherical accretion flow and
reveal the divergence of back reaction perturbations at both the event horizon and the inner
Cauchy horizon of the nearly extreme charged black hole. This singular behavior of back
reaction perturbations implies the violation of test fluid approximation for the stationary
spherical accretion onto extreme black holes.

The stationary accretion onto black hole is a solution with an independent of time $t$ and
radius $r$ influx of test fluid (or some field) with the energy density $\rho_\infty$ and
pressure $p_\infty$ far from the black hole.  We describe below the corrections to the
background spherically symmetric metric of the black hole due to the back reaction of
stationary accreting perfect fluid. The problem of a stationary accretion onto black hole is
self-consistent, if (i) the accreting fluid is lightweight, and (ii) the rate of black hole
mass growth is slow (the stationary limit). Two small parameters are needed to satisfy these
conditions. The first small parameter is a mass ratio of the inflowing gas in the volume with
the black hole gravitational radius, $m_{\rm gas}\sim\rho_\infty m^3$, to the black hole mass
$m$: $\rho_\infty m^3/m=\rho_\infty m^2\ll1$. With this small parameter, the test fluid
approximation in the background metric is valid inside the region  $r\ll R_{\rm
max}=m(\rho_\infty m^2)^{-1/3}$ around the black hole. The second small parameter is a slowness
of the black hole mass changing with respect to the hydrodynamical time, $\dot m/m\ll c_s/m$,
where $c_s$ is the sound velocity in the accreting gas. For example, the stationary spherical
accretion rate of the perfect fluid is \cite{michel,bde11}:
\begin{equation}
   \label{perfect}
    \dot m=4\pi A(\rho_\infty+p_\infty)m^2,
\end{equation}
where $A$ is a numerical constant, depending on the equation of state $p=p(\rho)$. This
constant $A$ is of the order of unity for relativistic fluids with $p\sim\rho$ (i.\,e, for
fluids with a relativistic sound velocity $c_s=O(1)$). Two small parameters become identical,
$\dot m\sim\rho_\infty m^2\ll1$, for the relativistic perfect fluids. The back reaction of
fluid on the background black hole metric in the considered stationary accretion limit may be
found by perturbation method due to existence of these two small parameters. Note that in the
stationary accretion limit the apparent horizon will coincide with the event horizon of the
accreting black hole. Quite the different approaches were used in
\cite{Shatsky99,TayHisAns00,WangHuang01,Shatsky07} for analysis of the back reaction on the
Reissner-Nordstr\"om black hole.

Below we describe the back reaction of the stationary accretion with a linear accuracy with
respect to the small parameter $\dot m=const\ll1$. We find analytically the corresponding back
reaction corrections to the black hole metric metric near the event horizon and near the inner
Cauchy horizon.

\section{Einstein equations}

A spherically symmetric metric may be written in the general form with two arbitrary functions
\cite{LL}:
\begin{equation}
 \label{spherical}
 ds^2=e^{\nu(t,r)} dt^2-e^{\lambda(t,r)}dr^2-r^2(d\theta^2+\sin^2\theta\,d\phi^2).
 \end{equation}
For application to the back reaction of the accreting matter on the charged black hole, we
define two metric functions, $f_0(t,r)$ and $f_1(t,r)$, and also two corresponding ``mass
functions''  (or mass parameters) $m_0(r,t)$ and $m_1(r,t)$:
\begin{eqnarray}
 \label{m0}
 e^{\nu(t,r)} &=& f_0(t,r)=1-\frac{2m_0(t,r)}{r}+\frac{e^2}{r^2}, \\
 e^{-\lambda(t,r)} &=& f_1(t,r)=1-\frac{2m_1(t,r)}{r}+\frac{e^2}{r^2},
 \label{m1}
\end{eqnarray}
where $e$ is the black hole electric charge. In the case of the Reissner-Nordstr\"om metric
(i.\;e., in the absence of accreting fluid), both mass functions coincide with the black hole
mass, $m_0=m_1=m$, and, respectively, $f_0=f_1=f$.

A spherically symmetric gravitational field in the general case is defined by the four Einstein
equations. Three of them are the differential equations of the first-order, and the fourth one
is of the second-order. These equations for metric (\ref{spherical}) have the following form
\cite{LL}:
\begin{eqnarray}
 \label{G01}
 8\pi T^1_0 &=& -e^{-\lambda}\frac{\dot\lambda}{r}, \\
 \label{G00}
 8\pi T^0_0 &=& -e^{-\lambda}\left(\frac{1}{r^2}-
 \frac{\lambda'}{r}\right)+\frac{1}{r^2}, \\
 \label{G11}
 8\pi T^1_1 &=& -e^{-\lambda}\left(\frac{1}{r^2}+
 \frac{\nu'}{r}\right)+\frac{1}{r^2}, \\
 \label{G22}
 8\pi T^2_2 &=& \frac{e^{-\nu}}{2}
\left[\ddot\lambda+\frac{\dot\lambda}{2}\left(\dot\lambda-\dot\nu\right)\right] \\
 &&-\frac{e^{-\lambda}}{2}
 \left[\nu''+(\nu'-\lambda')\left(\frac{\nu'}{2}+\frac{1}{x}\right)\right].
\end{eqnarray}
The corresponding components of the energy-momentum tensor for a perfect fluid are
\begin{eqnarray}
 \label{T01}
  T^1_0 &=&
 (\rho+p)u\sqrt{\frac{f_0}{f_1}(f_1+u^2)}, \\
 \label{T00}
 T^0_0 &=& \rho+(\rho+p)\frac{u^2}{f_1}+\frac{e^2}{8\pi r^4},  \\
 \label{T11}
 T^1_1 &=& -\left[(\rho+p)\frac{u^2}{f_1}+p\right]+\frac{e^2}{8\pi r^4}, \\
 \label{T22}
 T^2_2 &=& -p-\frac{e^2}{8\pi r^4},
 \end{eqnarray}
where $u=dr/ds$  is the radial component of the fluid 4-velocity, and, respectively, $\rho$ and
$p$ is an energy density and pressure of fluid in the comoving frame. Below it is supposed an
arbitrary equation of state $p=p(\rho)$, relating the fluid pressure and energy density. The
Bianchi identity holds true for Einstein equations, and so only three equations from four in
(\ref{T01})--(\ref{T22}) are independent. We choose (\ref{G01}), (\ref{G00}) and (\ref{G11})
for these independent equations.

Zero approximation in our approach corresponds to the stationary spherically symmetric inflow
of the test fluid in the background Reissner-Nordstr\"om metric. The corresponding solution
\cite{michel,carr74,bde11} defines the conserved radial flux of energy $\dot m$ and the radial
dependance for the 4-velocity component $u=dr/ds=u(r)$, for energy density $\rho=\rho(r)$ and
for pressure $p=p(\rho)=p(r)$. Respectively, this solution fixes all components of the
energy-momentum tensor $T^{ik}$. For self-consistency of the accretion problem in the
background metric, the radial influx of energy must be small, i.\;e. $\dot m\ll1$.

In the first approximation we will take into account the linear contributions with respect to
$\dot m\ll1$ to the energy momentum-tensor in Einstein equations. As a result, we find the
deviation of metric from the background one, i.\;e. the back reaction with a linear accuracy
with respect to the small parameter $\dot m$. It will be shown that the back reaction
corrections near the event horizon and inner Cauchy horizon of the black hole depend only on
the accretion rate $\dot m$ and do not depend on the equation of state $p=p(\rho)$ of the
accreting fluid. Formally, to find a back reaction modification of the black hole metric near
the horizons, it is needed to consider the space-time region, where both $f_0\ll1$ and
$f_1\ll1$.

\section{Back reaction in the Schwarzschild metric}

As the first step, we find the back reaction of accretion near the event horizon of the
Schwarzschild black hole. The first Einstein equation (\ref{G01}) defines the conserved radial
flux of anergy, i.\;e. the matter accretion rate,
\begin{equation}
 \label{Bernulli}
 \dot m=-4\pi r^2(\rho\!+\!p)u\sqrt{\frac{f_0}{f_1}\!(f_1\!+\!u^2)}
 =const.
 \end{equation}
The value of this flux (\ref{perfect}) is defined (in the zero approximation) from the
stationary solution of the test fluid accretion in the background Schwarzschild metric
\cite{michel,carr74,bde04,bde11}.  *** The radius of the modified event horizon, $r=r_+$, in
the considered stationary accretion limit is defined by the condition $f_0(r_+,t)=0$. The
Einstein equations are hyperbolic in general, and so we have at the horizon the second
condition, $f_1(r_+,t)=0$. ***

The first Einstein equation (\ref{G01}) in the linear approximation with respect to $\dot
m\ll1$ has a very simple form.
\begin{equation}
 \label{accretionflux2}
 \frac{\partial m_1}{\partial t}=\dot m.
 \end{equation}
This equation (in the considered stationary accretion limit with $\dot m\ll1$) alludes the
``factorization'' of the mass functions: $m_0(r,t)=m(t)\,\mu_0(r)$ and, respectively,
$m_1(r,t)=m(t)\,\mu_1(r)$. The dimensionless mass functions $\mu_0(r)$ and $\mu_1(r)$ are
defined so, that $\mu_0(r)=\mu_1(r)=1$ at $\dot m=0$. After substitution of this
``factorization'' ansatz in (\ref{accretionflux2}), we obtain
\begin{equation}
 \label{accretionflux3}
 \frac{d m(t)}{d t}\,\mu_1(r)=\dot m.
 \end{equation}
The r.h.s. of this equation is already linear with respect to $\dot m\ll1$. So, for the
dimensionless function $\mu_1(r)$ in this equation it is needed to use a zero approximation
with respect to $\dot m$, i.\;e. to put $\mu_1(r)=1$. In result, the partial differential
equation  (\ref{accretionflux2}) for the function $m_1(r,t)$ reduces to the ordinary
differential equation (\ref{accretionflux3}) for $m(t)$. The corresponding solution of this
reduced equation is
\begin{equation}
 \label{mt}
 m(t)=m(0)+\int_0^t\dot m(t') dt'.
\end{equation}
Here $m(0)$ is a black hole mass at some initial moment $t=0$ and $m(t)$ is a current value of
black hole mass. The black hole mass $m(t)$ is slowly growing due to a slow accretion rate
$\dot m\ll1$. This is the only one equation in our analysis, where the temporal dependance of
the accretion rate $\dot m(t)$ must be taken into account. In all other equations the accretion
rate $\dot m(t)$ is only a small constant parameter.

At this step we may find the function $\mu_1(r)$ with the help of the second Einstein equation
(\ref{G00}). After factorization of the mass functions $m_0l$ and $m_1$, it is useful in the
following to use the dimensionless radial variable $x=r/m(t)$ with $m(t)$ from (\ref{mt}). The
second Einstein equation (\ref{G00}) is written now in the form
\begin{equation}
 \label{flux2}
 \frac{d\mu_1}{dx}=4\pi x^2\left[\rho+(\rho+p)\frac{u^2}{f_1}\right].
\end{equation}
A combined solution of equations (\ref{Bernulli}) and (\ref{flux2}) with the using of accretion
solution for fluid with equation of state $p=p(\rho)$, defines the requested function $\mu_1$.
Near the event horizon, where $f_0\ll1$ and $f_1\ll1$, from (\ref{Bernulli}) and (\ref{flux2})
we obtain
\begin{equation}
 \label{flux2m22}
 \frac{d\mu_1}{dx}\approx\frac{4\pi x^2(\rho+p)u^2}{f_1}
 \approx\frac{2\dot m}{x-2\mu_1(x)}.
\end{equation}
By using a new variable $\delta(x)=x-2\mu_1(x)\ll1$, we find the approximate solution of
equation (\ref{flux2m22}) near the event horizon $x_+$, defined by the condition
$\delta(x_+)=x_+-2\mu_1(x_+)=0$:
\begin{equation}
 \label{mu1}
 \mu_1(\delta)\approx\mu_+ +2\dot m\log|\frac{\delta}{4\dot m}-1|,
\end{equation}
where
\begin{equation}
 \label{mu+}
 \mu_{+}=\mu_1(x_+)\approx1+2\dot m\log|\dot m|.
\end{equation}
is the value of mass function $\mu_1(x)$ at the event horizon radius $x_+$, modified by the
back reaction. Solution (\ref{mu1}) for the inverse function $x=x(\mu_1)$ is written in form:
\begin{equation}
 \label{mdelta2}
 x(\mu_1)\approx 2\mu_++4\dot m\left(\!1+\frac{\mu_1-\mu_+}{2\dot m}
 -\exp{\frac{\mu_1-\mu_+}{2\dot m}}\!\right)\!,
\end{equation}
where $x_+=x(\mu_+)= 2\mu_+$. Solution (\ref{mu1}) for the Schwarzschild metric, modified by
the back reaction of accreted matter is shown in Fig.~1. It is important to note that this
solution is valid only in the narrow region $8\dot m=|\delta_{\rm min}|<|\delta|\ll1$ around
the modified event horizon $x_+$. The used linear approximation with respect to $\dot m$ would
be insufficient for calculation of the corresponding mass function at $|\delta|<|\delta_{\rm
min}|$ (inside the filled box in Fig.~1).

During integration of equation (\ref{flux2}) we retained only the leading term with $f_1\ll1$
in the denominator, providing the major contribution to the solution  $\sim\dot m\log|\dot m|$,
and neglected all contributions to the solution of the order of $\dot m$. Formally, we supposed
that distribution of fluid around the black hole is a finite sphere of radius $X_0$, satisfying
the condition $1\ll X_0\ll X_{\rm max}=(\rho_\infty m^2)^{-1/3}$. We neglect the contribution
to solution of equation (\ref{flux2}) from the term $4\pi x^2\rho$, related with mass of the
accreting gas inside the sphere of radius $X_0$, and also the contribution from the leading
term at the upper integration limit $\sim\dot m/X_0$, which is the gravitational ``mass
defect''. Finally, we neglect in the integration constant the contribution from the boundary
condition, which is also $\sim\dot m$.
\begin{figure}
\begin{center}
    \includegraphics[width=0.48\textwidth]{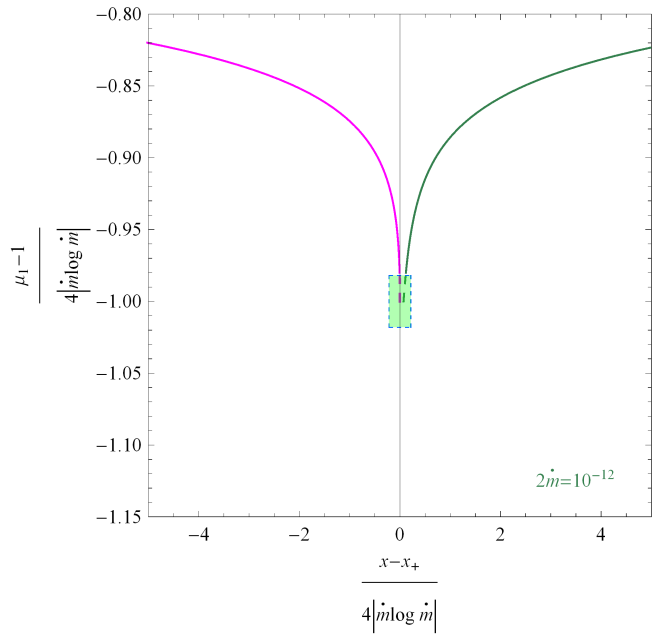}
 \label{ultrah}
 \end{center}
\caption{The mass function $\mu_1(x)$ from equation (\ref{mu1}) near the event horizon of black
hole $x_+$, modified by back reaction, in the linear approximation with respect to $\dot
m\ll1$. Inside the filled box the used linear approximation is insufficient for determination
of the mass function $\mu_1(x)$.}
\end{figure}

The third Einstein equation (\ref{G11}) for the considered perfect fluid has the form
\begin{equation}
 \label{3}
 \frac{f_1}{f_0}\mu_0'\!+\!\frac{1}{x}\!\left(\mu_1-\mu_0\frac{f_1}{f_0}\right)
 =-4\pi x^2\!\left[(\rho+p)\frac{u^2}{f_1}+p\right]\!.
\end{equation}
Near the event horizon this equation is written as
\begin{equation}
 \label{3+}
\mu_0'\approx-4\pi x^2(\rho+p)\frac{u^2}{f_1}\approx-\mu_1'.
\end{equation}
Here it is taken into account that a ratio $f_0/f_1$ near the horizon equals to its background
value, $(f_0/f_1)_+=1$, at the linear approximation with respect to $\dot m\ll1$. Solution of
equation (\ref{3+}) near the event horizon, where $\delta=x-2\mu_1(x)\ll1$, with the help of
(\ref{mu1}) and (\ref{3+}) is
\begin{equation}
 \label{mu0}
 \mu_0(\delta)\approx\mu_+ - 2\dot m\log|1-\frac{\delta}{4\dot m}|.
\end{equation}
By comparing (\ref{mu1}) and (\ref{mu0}), we see that near the horizon, at $|\delta|\ll1$, it
is satisfied the condition
\begin{equation}
 \label{mu01}
 \mu_0(\delta)+\mu_1(\delta)\approx2\mu_+.
\end{equation}
To calculate the corresponding back reaction in the case of the Reissner-Nordstr\"om black hole
(\ref{spherical}) -- (\ref{m1}) we define the extreme parameter of the black hole
$\epsilon=\sqrt{1-e^2/m^2}$ and a new variable $\delta_{\pm}=x-[\mu_1(x)\pm
\sqrt{\mu_1(x)^2-1+\epsilon^2}]$.

\section{Back reaction in the Reissner-Nordstr\"om metric}

Quite similar to the Schwarzschild case, near the event horizon $x_+$ and the inner Cauchy
horizon $x_-$ of the Reissner-Nordstr\"om black hole we find:
\begin{eqnarray}
 \mu_1(\delta_\pm)&\approx&1\pm\frac{(1\!\pm\epsilon)^2}{2\epsilon}\,\dot m
 \log{\left|\frac{(1\!\pm\epsilon)^3}{2\epsilon^2}\,\dot m-\delta_{\pm}\right|},  \label{muRN1}
 \\
 \mu_0(\delta_\pm)&\approx&2\mu_\pm-\mu_1(\delta_\pm),
  \label{muRN0}
\end{eqnarray}
where
\begin{equation}
 \label{mupm}
 \mu_\pm = 1\pm\frac{(1\!\pm\epsilon)^2}{2\epsilon}\,\dot m
 \log{\left|\frac{(1\!\pm\epsilon)^3}{2\epsilon^2}\,\dot m\right|}
 \end{equation}
This solution is valid only at $[(1\!\pm\epsilon)^3\!/\epsilon^2]\dot m <|\delta_\pm|\ll1$ and
$\delta_- > 0$. The value of $\mu_1(x)$ at the modified event horizon $x=x_+$ and at the inner
Cauchy horizon $x=x_-$ corresponds formally to $\delta_{\pm}=0$ in (\ref{muRN1}). The
corresponding radii of horizons are
\begin{eqnarray}
 \label{horizonRN}
 x_{\pm}  &=&\mu_1(x_{\pm})\pm\sqrt{\mu_1(x_{\pm})^2-(1-\epsilon^2)} \\
 &\approx& \!(1\pm\epsilon)\!\left[1
 \pm\frac{1}{2}\!\left(\!\frac{1\pm\epsilon}{\epsilon}\!\right)^{\!2}\!\dot m
 \log{\left|\frac{(1\pm\epsilon)^3}{\epsilon^2}\dot m\right|}\right]\!.
 \nonumber
 \end{eqnarray}
We uphold here the major back reaction term $\propto\dot m\log|\dot m|$ and neglect the much
smaller term $\propto\dot m$.

\section{Conclusions}
\begin{figure}[t]
\begin{center}
    \includegraphics[width=0.48\textwidth]{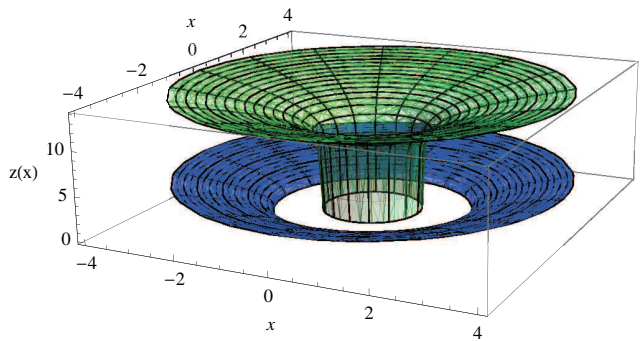}
 \end{center}
\caption{The shallow funnel is the embedding diagram for the Shwarzschild BH (Flamm's
paraboloid) and the deep funnel is the corresponding embedding diagram for the near extreme
Reissner-Nordsr\"om BH. The height of the funnel tends to infinity at   $\epsilon\to 0$
according to (\ref{local}).} \label{embedd}
\end{figure}

From (\ref{muRN1}) -- (\ref{horizonRN}) it follows directly, that the test fluid approximation
{\sl is violated} due to the back reaction of the accreted fluid in the extreme black hole
limit $\epsilon\to0$. Namely, the corresponding corrections to the radius of the black hole
event horizon and the inner Cauchy horizon diverge at any arbitrarily small accretion rate
$\dot m$, if $\epsilon\to0$. This behavior is in agreement with the cosmic censorship
conjecture \cite{penrose69} and with the third law of black hole thermodynamics
\cite{bardeen73}: the extreme state is unattainable in the finite processes or, in other words,
it is impossible in practice to transform the black hole into the naked singularity. A similar
conclusion was derived recently in \cite{japan11} by demonstration that the
Reissner-Nordstr\"{o}m black hole can never be overcharged to the naked singularity via the
absorption of a charged particle.

Note, that the test fluid approximation is violated for the stationary accretion of ultra-hard
fluid with $p=\rho$ at the event horizon of the extreme Kerr-Newman black hole even without the
back reaction corrections \cite{PetShapTeu,bcde08,bde11}. Violation of the test fluid
approximation for the accretion onto extreme black holes is also in accordance with the absence
of solutions for stationary accretion onto the naked singularities \cite{bcde08,bde11}. To
resolve the problem of back reaction of accreting matter on the extreme black hole it is
requested to find a solution of Einstein equations beyond the perturbation level.

A physical reason for the divergence of back reaction corrections in the extreme case is in the
infinite stretching of a local distance $l(x)$ to the event horizon $r_+$ (and, analogously, to
the Cauchy horizon $r_-$) at $\epsilon\to 0$:
\begin{eqnarray}
 \label{local}
 \frac{l(x)}{m} &=& \frac{1}{m}\int_{r_+}^r\!f^{-1/2}\, dr =
 \int_{x_+}^x\frac{x'dx'}{\sqrt{x'^2-2x'+e^2}} \\
 &=&\!xf^{1/2}\!+\log\!\left[1\!+\!
 \frac{\sqrt{x\!-\!x_+}}{\epsilon}(\sqrt{x\!-\!x_+}+\sqrt{x\!-\!x_-})\right].
 \nonumber
\end{eqnarray}
See in Fig.~\ref{embedd} the corresponding embedding diagram for the Reissner-Nordsr\"om BH,
constructed from the relation $dz^2+dx^2= f^{-1}dx^2$ with $f$ from (\ref{m0}) and (\ref{m1}).
The space near horizons is loaded with a finite energy density $\rho(x)$ of the accreting fluid
along the local distance $l(x)$, providing contribution to the mass functions $m_0(x)$ and
$m_1(x)$.

Due to the infinite stretching of $l(x)$ in the extreme case $\epsilon=0$, the total mass of
accreting gas and the corresponding mass functions $\mu_0(x)$ and $\mu_1(x)$ in (\ref{muRN1})
and  (\ref{muRN0}) are diverging near the extreme black hole. Quite a similar infinite
stretching of $l(x)$ is also a characteristic feature of the extreme Kerr black hole
\cite{bpt72}. Therefore, the described violation of the test fluid approximation for the
stationary accretion might be crucial both for the extremely charged and extremely rotating
black holes.

\acknowledgments We acknowledge E. Babichev and V. Berezin for helpful discussions. This
research was supported in part by the Russian Foundation for Basic Research through Grant No.
10-02-00635.

\end{document}